\documentclass[usenatbib,onecolumn]{mnras}
\usepackage{amsmath}
\usepackage{epsfig,color}
\usepackage{dcolumn}
\usepackage[utf8]{inputenc}
\usepackage{graphics,graphicx}
\usepackage{epstopdf}
\usepackage{float}
\usepackage[normalem]{ulem}

\graphicspath{ {eps/} }

\usepackage{url}
\usepackage{color}

\usepackage{braket}
\usepackage{lscape}

\newcommand{\cm}{cm$^{-1}$}

\newcommand{\ai}{\textit{ab initio}}

\title[ExoMol line lists XXV: SiS]{ExoMol line lists XXV: A hot line list
for silicon sulphide, SiS}
\date{\today}

\author[Upadhyay et al]{\large Apoorva Upadhyay, Eamon K. Conway, 
Jonathan Tennyson\thanks{Email: j.tennyson@ucl.ac.uk} and Sergei N. Yurchenko \\
Department of Physics and Astronomy, University College London, Gower
Street, London, UK, WC1E 6BT } 
\date{Accepted XXXX. Received XXXX; in original form XXXX}
\pagerange{\pageref{firstpage}--\pageref{lastpage}} \pubyear{2017}

\begin{document}

\label{firstpage}

\maketitle

\begin{abstract}
  SiS has long been observed in the circumstellar medium of the carbon-rich star IRC+10216 CW Leo.   Comprehensive and accurate  
  rotation-vibrational line lists and partition functions are
computed for 12 isotopologues of silicon sulphide
  ($^{28}$Si$^{32}$S, $^{28}$Si$^{34}$S, $^{29}$Si$^{32}$S, $^{28}$Si$^{33}$S, $^{30}$Si$^{32}$S,
  $^{29}$Si$^{34}$S, $^{30}$Si$^{34}$S, $^{28}$Si$^{36}$S, $^{29}$Si$^{33}$S, $^{29}$Si$^{36}$S,
$^{30}$Si$^{33}$S, $^{30}$Si$^{36}$S)
in its ground (X $^1\Sigma^+$)
  electronic state. The calculations
  employ an existing spectroscopically accurate potential energy curve (PEC)
  derived from experimental measurements and a newly-computed  {\it ab initio} dipole
  moment curve (DMC). The $^{28}$Si$^{32}$S  line list includes 10~104	states
and 91~715 transitions. 
These line lists are available from the ExoMol website
(\url{www.exomol.com}) and the CDS database.


\end{abstract}


\section{Introduction}

Silicon sulphide is well known in space.  The first detections of SiS
were in the microwave region by \citet{75MoGiPa.SiS}, emitting from
the IRC+10216 molecular envelope, who deduced that SiS was in greater
abundance than SiO. This was in line with \citet{73Tsuji} who had
earlier suggested that to detect SiS in circumstellar envelopes, the
abundance ratio of carbon to oxygen needs to be much greater than
unity, because the abundance of SiS is a function of the [C]/[O]
abundance ratio.  At larger distances away from the star, SiS
molecules condense onto dust grains.  Silicon from the dust grains is
thought to be released into gas phase via shock waves produced due to
the pulsation of the star.  In regions radially further still, the
ultra-violet (UV) radiation in the interstellar medium dissociates the
molecules, hence the abundance of SiS is expected to be low in these
regions \citep{15VeCeQu.SiS}.

\citet{75MoGiPa.SiS} concluded that the observed radio-frequency lines of
molecules containing elements such as Si and S could provide
information on the nuclear processes occurring in stars which are in
their post main sequence (PMS) phase. Subsequent detections of
many rare isotopologues of SiS were reported by \citet{84JoAnEl.SiS},
\citet{84ZiClSa.SiS}, \citet{88KaGoCe.SiS}, \citet{00CeGuKa.SiS} and
\citet{04MaOtHe.SiS}.  Maser emission from IRC+10216 was reported
by  \citet{83HeMaMo.SiS}. They observed $J = 1 - 0$ transitions of SiS in
the vibrational state $v = 0, 1$ and 2 at frequencies near 18 GHz. 
According to \citet{83HeMaMo.SiS}, there is a population inversion in
the $J = 1 - 0$ transition in the ground vibrational state which is
responsible for the maser emission.  \citet{06ExAgTe.SiS}
reported the first detections of SiS maser emission from $J
= 15 - 14$, $J = 14 - 13$ and $J = 11 - 10$ transitions in the ground
vibrational state from IRC+10216.   First reports of vibrationally excited
SiS in IRC+10216 came from \citet{87Turner.SiS} who
detected transitions within
the vibrational excited $v = 1$ state. \citet{87Turner.SiS} concluded that the
emission arises from the inner region of the circumstellar
envelope which has a temperatures greater than 600 K. More recently
\citet{15VeCeQu.SiS} observed 
rotational lines of SiS in high vibrational states. Using the Atacama Large
Millimetre Array (ALMA) radio telescopes they report detections of
rotational emission lines for excited  vibrational states as high as $v =
7$; transitions for
other isotopologues including $^{29}$Si$^{32}$S, $^{30}$Si$^{32}$S, 
$^{28}$Si$^{33}$S, $^{28}$Si$^{34}$S, $^{29}$Si$^{33}$S and
$^{29}$Si$^{34}$S in high vibrational states were also observed.
Observations of 24 rotation-vibrational lines of SiS from
IRC+10216 are reported by \citet{04BoKeJe.SiS} who estimated
the rotational excitation temperature to be $704 \pm  85$ K.
Considerable work continues on observing SiS spectra; for example,
recently \citet{17DaVaDe.SiS} observed  several  lines  of  SiS and other S-containing species  in
a  diverse  sample  of  20  AGB  stars,  including  7  M-type  stars,
5 S-type stars, and 8 carbon stars.

The ExoMol project  aims at providing line lists of
spectroscopic transitions for key molecular species which are likely to be
important in the atmospheres of extrasolar planets and cool stars
\citep{jt528,jt631}.  This is essential for the continued exploration of newly
discovered astrophysical objects such as exoplanets, for which there is an
increasing desire to characterise their atmospheric compositions.  The
methodology of the line list production for diatomics is discussed by
\cite{jt626}. ExoMol has already provided rotation-vibration line lists for
several silicon-containing molecules: SiO \citep{jt563},
SiH$_4$ \citep{jt701} and SiH \citep{jt711}, and for several sulphur-containing molecules:
CS \citep{jt615}, PS \citep{jt703}, H$_2$S \citep{jt640}, SO$_2$ \citep{jt635} and SO$_3$ \citep{jt641},
as well as most recently SH and SN \citep{jt725}.
Given the astronomical importance of SiS, we present 
line lists for the 12 stable isotopologues of SiS applicable for temperatures up to 5000~K.

The following section discusses, respectively, the avaiable experimental
and theoretical data for the SiS molecule. Section~\ref{s:method} describes our methodology. 
Section~\ref{s:linelists} presents our results and compares with previous data.
Finally Section \ref{s:concl} briefly presents our conclusions.

\section{Previous laboratory studies}
\subsection{Experimental data}

\citet{38BaJexx.SiS} first observed D $^1\Pi$ -- X $^1\Sigma^+$ SiS
band in the ultraviolet (UV) region 2500 to 6500 \AA.  Later the
E~$^1\Sigma^+$-- X $^1\Sigma^+$ band system was observed by
\citet{46VaBaxx.SiS} in absorption at temperatures of about 1000 $^\circ$C.
These bands were further analysed by \citet{46Barrow.SiS},
\citet{61BaLaDe.SiS}, \citet{75BrCoDu.SiS}, \citet{76BrCoDu.SiS} and
\citet{85LaShGo.SiS}.  \citet{80Linton} observed chemiluminescent
following the formation of SiS molecules in the reaction of Si atoms
with OCS. The spectra showed two main bands in the region 350 – 400 nm
and 385 – 600 nm attributed to transitions within the $^3\Pi$ -- X
$^1\Sigma^+$ and $^3\Sigma^+$ -- X $^1\Sigma^+$ systems.

The rotational spectrum of SiS was measured by \citet{69HoLoTi.SiS},
\citet{70HoLoTi.SiS} and  \citet{72TiReHo.SiS}. The permanent dipole
moment in the ground state was determined by Stark effect
measurements  to be $\mu = 1.74 \pm 0.07$ D  by \citet{69MuCuxx.SiS}
and also by \citet{69HoLoTi.SiS}. Isotopic effects
on the rotational spectrum of SiS were investigated by \citet{72TiReHo.SiS}
who obtained Born-Oppenheimer breakdown corrections.

More recently \citet{07MuMcBi.SiS} observed 300 pure rotational
transitions of SiS and its 12 stable isotopic species in the
vibrational ground state and vibrationally excited states.  Rotational
transitions were observed for the isotope of least abundance
$^{30}$Si$^{36}$S isotopologue in the ground vibrational state, as
well as rotational transitions in $v = 1$ for $^{28}$Si$^{32}$S.
\citet{90FrEnBe.SiS} recorded the rotation-vibration spectrum of SiS
at 13 $\mu$m (750 \cm) using Fourier transform emission spectroscopy;
they recorded seven bands for the parent isotopologue of SiS and three
bands for each of the rarer isotopologues. \citet{90BiJoxx.SiS} also
measured the rovibrational spectrum for four isotopologues of SiS
($^{28}$Si$^{32}$S, $^{28}$Si$^{34}$S,
$^{29}$Si$^{32}$S,$^{30}$Si$^{32}$S) in the ground electronic state.
Further experimental data for the vibrational energy levels of the
ground electronic state of SiS is provided by \citet{65NaSiRa.SiS}.

\subsection{Theoretical data}

Several \ai\ studies have been carried out on SiS starting with 
\cite{81RoLeCo.SiS} who computed spectroscopic parameters of electronic
states of SiS.  Potential energy curves (PECs) for the ground
electronic and various excited states have been calculated by several
authors. These include finite difference Hartree-Fock (HF)
calculations on the ground electronic state by \citet{89MuLaxx.SiS}.
\citet{02ChChDa.SiS} 
computed PECs for a number of lower electronic states of SiS using
configuration interaction calculation with
relativistic effective core potentials. 

\citet{92CoHaxx.SiS} determined an empirical X $^1\Sigma^+$ state PEC 
using observed rotational –
vibrational and pure rotational transition line positions. 
This PEC, which includes Born-Oppenheimer breakdown (BOB) corrections, is accurate to within
experimental error. Coxon and Hajigeorgiou's PEC and BOB corrections
are used in this work. See section 3.2 for more details.

\citet{88LiMoZh.SiS} focused their calculations on the X $^1\Sigma^+$
electronic ground state of SiS and used multi reference configuration
interaction (MRCI) level calculations to compute the dissociation
energy, the equilibrium bond length and a dipole moment curve
(DMC).  \citet{88LiMoZh.SiS} obtained their best value of $\mu$ = 1.57 D for
the permanent dipole moment, compared to the measured value $\mu =
1.74 \pm 0.07$ D; they suggested that the addition of diffuse functions to
the basis sets should account for the discrepancy between the
calculated and experimental value. Subsequently, \citet{93HuMiSe.SiS}
also calculated the ground electronic DMC of SiS at the SCF
(self-consistent field) level and obtained a value of $\mu$ = 2.170 D.
\citet{00MaMaXe.SiS} performed coupled cluster (CCSD(T)) and finite
field many body perturbation theory calculations to obtain a permanent
dipole moment value close to that of \citet{88LiMoZh.SiS}, $\mu$ = 1.556
D.  \citet{11ShXiZh.SiS} provided us (private communication, 2017) a
DMC computed at the MRCI level with a large aug-cc-pV6Z basis set
which gives $\mu$ = 1.611 D.  \citet{87PiTiCh.SiS} provided a semi empirical dipole moment function which
they used to estimate dipole matrix elements for vibration-rotational transitions. 
Given the variation in theoretical dipoles and the lack of agreement with the measured values, we
compute our own ab initio DMC, see section 3.1.

\section{Method}
\label{s:method}

The general procedure adopted here is similar to that used by us for
other closed-shell diatomics such as SiO \citep{jt563}, PN \citep{jt590}
and CS \citep{jt615}.  The nuclear motion problem was solved using the
program Level \citep{Level}. As input we used the
spectroscopically-determined PEC of \citet{92CoHaxx.SiS}, with minor
adjustments caused by discretization of the PEC as described below, and an \ai\ DMC presented below.

\subsection{Dipole moment curve}\label{sec:dipole}

Initially we tested the calculated DMC of \citet{14LiZhLi.SiS}. However, when we compared these
values to those given on the CDMS (The Cologne Database for Molecular
Spectroscopy) database \citep{cdms,07MuMcBi.SiS}, we found large discrepancies in the values of the Einstein A coefficients so decided to
calculate our own DMC.

{\it Ab initio} calculations
of the DMC were performed using MOLPRO \citep{12WeKnKn.methods} at the
CCSD(T) level with an aug-cc-pV5Z basis set for 128 points between
0.9 and 3.2 \AA. The dipoles were computed using the finite field approach
(see \citet{jt475}) and stable results required using a low 
perturbing electric field strength of 0.00005 atomic units. 

\begin{figure}
\centering
\includegraphics[scale=0.6]{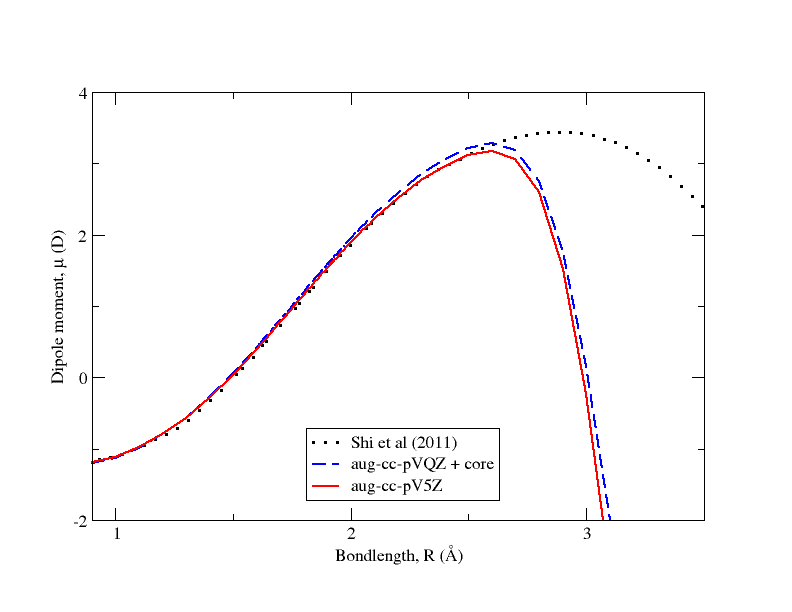}
\caption{{\it Ab initio} dipole moment curves as computed by \citet{11ShXiZh.SiS} and this work.}
\label{f:DMF}
\end{figure}

Figure \ref{f:DMF} compares {\it ab initio} DMCs. Except for the
curve provided by \citet{11ShXiZh.SiS}, all other dipole curves have been
calculated as part of this study using the CCSD(T) method. 
Our CCSD(T) DMCs appear to drop too rapidly at large bondlength.
This behaviour appears to be a feature of CCSD(T) DMCs \citep{jt573}.
However, tests showed that our results are not sensitive to the DMC beyond
$R=2.8$~\AA. Conversely our aug-cc-pV5Z  DMC is smooth at bondlengths about
equilibrium but that due to \citet{11ShXiZh.SiS} is computed
at fewer points and is then less smooth when the points are
used directly in the nuclear motion calculation. This lack of smoothness
in the DMC leads to unphysical intensities \citep{highv}.
Our aug-cc-pV5Z dipole points are included in the supplementary
material to this article as part of the input to Level.

Einstein A coefficients were calculated using Level and our CCSD(T)
aug-cc-pV5Z DMC.  Our computed value of the equilibrium dipole
moment is $\mu$ = 1.70 D whereas \citet{11ShXiZh.SiS} calculate this
to be $\mu$ = 1.61 D; the experimental value for the $v = 0$ dipole
moment is $\mu = 1.74 \pm 0.07$ D. This higher value for our
dipole gave better agreement with results of CDMS for pure rotational
spectra, see below. Conversely all reliable DMCs considered give
rather similar slopes in the region of equilibrium. This leads
to a somewhat larger transition dipoles for the vibrational fundamental
than the one assumed by CDMS, who use transition dipole values based on the semi-empirical estimates of \citet{87PiTiCh.SiS}.

\subsection{Potential Energy Curve}

A very accurate PEC was derived by \citet{92CoHaxx.SiS} by fitting to
spectroscopic data. The fit employed microwave and infrared data on four isotopologues 
($^{28}$Si$^{32}$S, $^{28}$Si$^{34}$S,$^{29}$Si$^{32}$S,$^{30}$Si$^{32}$S) from \citet{72TiReHo.SiS}, \citet{90BiJoxx.SiS} and \citet{90FrEnBe.SiS}. 
The resulting PEC is valid for all isotopologues of SiS.  Their Born-Oppenheimer SiS potential energy
function takes the functional form of a $\beta$-variable Morse
potential \citep{91CoHaxx.diatom} with additional atomic mass dependent BOB terms added. The effective PEC takes the form
\begin{equation}\label{eq:bobterms}
U^{eff}_{SiS}(R) = U^{BO}_{SiS}(R) + \frac{1}{M_{Si}} \sum\limits^{3}_{i=1} u^{Si}_i (R-R_{e})^i + \frac{1}{M_{S}} \sum\limits^{2}_{i=1} u^{S}_i (R-R_{e})^i
\end{equation}
where the last two terms give the functional forms of the J-independent BOB corrections for SiS. 
$M_{Si}$ and $M_{S}$ are the atomic masses 
of the isotopes of Si and S respectively. Initially \citet{90BiJoxx.SiS} tried to invert their 
measured line positions data to a Born-Oppenhiemer potential as a Dunham expansion but were unable 
to fit high J line positions ($J \geq 100$), which had to be excluded from the fit. 
\citet{92CoHaxx.SiS} are able to include this data in the final fit using their model.

\citet{92CoHaxx.SiS} showed that a small number of fitted parameters
were able to represent the entire spectroscopic dataset to within the
measurement accuracies of about 0.001 \cm\ for the measurements of \citet{90BiJoxx.SiS} 
and 0.0001 \cm\ for measurements of the strongest lines from \citet{90FrEnBe.SiS}.

However, the functional form of the $\beta$-variable Morse
potential is not one of those included in Level.
Therefore, the expansion parameters of the PEC given by 
\citet{92CoHaxx.SiS} were used to generate data points of the PEC that
could be directly input into Level. The PEC was generated on a grid
of 0.001 \AA\ from 1.0 to 3.0 \AA. These points are included in the
sample Level input given in the supplementary data. As the BOB term
is isotopologue dependent, it was necessary to generate a new grid of effective
PEC points for each isotopologue.
Tests for $^{28}$Si$^{32}$S showed that the results, and in particular
the number of vibrational state obtained, were insensitive to extending
this range.

.
\subsection{Nuclear motion calculations}

Nuclear motion calculations were performed using the program Level
\citep{Level}. All vibrational states were considered for the given
PEC and isotopologue. The eigenvalues were calculated in Level 
using an eigenvalue convergence parameter (EPS) value set to $10^{-8}$ \cm.
 Table~\ref{t:Nair} compares our results
for the vibrational term values (ie states with $J=0$) with the
measurements of \citet{65NaSiRa.SiS} and the calculations of
\citet{92CoHaxx.SiS}. It can be seen that both calculations agree
equally well with the observation and that there is a slight shift,
about 0.003 \cm\ between the two theoretical calculations.

The discretization of the PEC 
and minor changes in the fundamental constants use probably account for this small shift. \citet{92CoHaxx.SiS} remark that in order 
to exactly reproduce their vibrational term values the constant $h/8\pi^{2}c$ should be set to a value 
of 16.8576314 $amu$ $\AA^2$  \cm. However in Level this constant is fixed at a value of 16.857629206 $amu$ $\AA^2$ \cm. 
Given that this shift is almost uniform and we are interested in precise transition frequencies rather
than energy levels, this shift was not considered important.

\begin{table}
\caption{Comparison of calculated parent isotopologue $^{28}$Si$^{32}$S vibrational energy
 levels, in \cm,
from  \citet{92CoHaxx.SiS} (Coxon) and this work with
the empirical values (Obs) of \citet{65NaSiRa.SiS}} 
\label{t:Nair}
\begin{tabular}{rrrrrr}
\hline\hline
\multicolumn{1}{c}{$v$}	&	\multicolumn{1}{c}{Obs.}	&
\multicolumn{1}{c}{Coxon}&        \multicolumn{1}{c}{This Work}	&
\multicolumn{1}{c}{obs$-$calc}	&\multicolumn{1}{c}{Coxon$-$This work}	
\\
\hline
0  &374.2   &374.2077  &374.2114 &0.0   &-0.0036   \\

1  &1119.0  &1118.6843  &1118.6877&0.3   &-0.0035   \\

2  &1858.3  & 1857.9975 &1858.0009&0.3   &-0.0033   \\

3  &2592.0  & 2592.1534 &2592.1566&-0.2  &-0.0032   \\

4  &3321.8  & 3321.1575 &3321.1606&0.6   &-0.0031   \\

5  &4045.9  & 4045.0155 &4045.0185&0.9   &-0.0030   \\

6  &4763.5  & 4763.7328 &4763.7356&-0.2  &-0.0028   \\

7  &5476.9  & 5477.3144 &5477.3171&-0.4  &-0.0027   \\

8  &6181.6  & 6185.7655 &6185.7682&-4.2  &-0.0027   \\

9  &6886.8  & 6889.0909 &6889.0936&-2.3  &-0.0027   \\

10 &7587.5  & 7587.2953 &7587.2980&0.2   &-0.0027   \\
\hline\hline
\end{tabular}
\end{table}

\begin{table}
\caption{Comparison of vibrational energy levels, in \cm,
from  \citet{92CoHaxx.SiS} (Coxon) and this work for isotopically subsituted SiS.} 
\label{t:iso}
\begin{tabular}{llrrc}
\hline\hline
&\multicolumn{1}{c}{$v$}		&
\multicolumn{1}{c}{Coxon}&        \multicolumn{1}{c}{This Work}	&
\multicolumn{1}{c}{Coxon$-$This work}	
\\
\hline

$^{28}$Si$^{34}$S   
&0&  369.0503&  369.0538& -0.0035 \\
&1& 1103.3198& 1103.3231& -0.0034 \\
&2& 1832.5673& 1832.5706& -0.0032  \\
&3& 2556.7986& 2556.8017& -0.0031  \\
&4& 3276.0191& 3276.0221& -0.0030  \\
                                   \\
$^{29}$Si$^{32}$S     
&0&  370.7550&  370.7586& -0.0036  \\
&1& 1108.3985& 1108.4019& -0.0034  \\
&2& 1840.9736& 1840.9769& -0.0033  \\
&3& 2568.4860& 2568.4891& -0.0031  \\
&4& 3290.9413& 3290.9443& -0.0030  \\
                                 \\
$^{30}$Si$^{32}$S     
&0&  367.5104&  367.5139& -0.0035 \\
&1& 1098.7321& 1098.7354& -0.0033\\
&2& 1824.9737& 1824.9769& -0.0032\\
&3& 2546.2409& 2546.2440& -0.0031  \\
&4& 3262.5390& 3262.5420& -0.0030 \\
\hline\hline
\end{tabular}
\end{table}

Table~\ref{t:iso} compares predicted vibrational band origins
for the three most important isotopically substituted SiS molecules with
the results of \citet{92CoHaxx.SiS}. Again the results show a small,
almost uniform, systematic shift in the region of 0.003 \cm. Again,
this difference
is probably not significant.

\section{Line lists}\label{s:linelists}

\subsection{Partition Function}
\begin{table}
\caption{Statistics for line lists for the twelve isotopologues of SiS considered in this work.}
\begin{tabular}{lrrrr}
\hline\hline
Isotopologue&	 $v_{\rm max}$	& $J_{\rm max}$ &	Number of energies&	Number of lines  \\
\hline
$^{28}$Si$^{32}$S&	42&	257&	10104&	91715  \\
$^{28}$Si$^{34}$S&	42&	257&	10251&	94282  \\
$^{29}$Si$^{32}$S&	42&	257&	10204&	92003  \\
$^{28}$Si$^{33}$S&	42&	257&	10182&	91941  \\
$^{28}$Si$^{36}$S&	43&	257&	10423&	94751  \\
$^{30}$Si$^{34}$S&	43&	257&	10487&	94932  \\
$^{29}$Si$^{34}$S&	43&	257&	10387&	94658  \\
$^{29}$Si$^{33}$S&	42&	257&	10277&	94378  \\
$^{30}$Si$^{33}$S&	43&	257&	10411&	94709  \\
$^{29}$Si$^{36}$S&	43&	257&	10528&	95036  \\
$^{30}$Si$^{36}$S&	44&	257&	10663&	95294  \\
$^{30}$Si$^{32}$S&	43&	257&	10316&	94501  \\
                 
\hline\hline           
\end{tabular}
\label{t:stats}
\end{table}

Level was used to compute all  bound rotation-vibration
states of each of the 12
isotopologues considered, see summary in Table~\ref{t:stats}. 
Partition functions were then calculated
by direct summation of all energy levels. Contributions from
quasi-bound or electronically excited states were ignored.
Since the nuclear
spin degeneracy of both $^{28}$Si and $^{32}$S is zero, the
nuclear spin degeneracy factor for  $^{28}$Si$^{32}$S is unity
which is the value adopted by all conventions. For the
other isotopologues we follow the convention adopted by HITRAN \citep{jt692}
and use full integer weights given by $(2I({\rm Si})+1)((2I({\rm S}) +1)$,
where $I(X)$ is nuclear spin of species $X$.

Table~\ref{t:pf} compares our partition function for $^{28}$Si$^{32}$S
with previous compilations. The agreement is excellent.
\citet{16BaCoxx.partfunc} calculate their partition function values
from spectroscopic constants compiled by \citet{79HeHuxx.book} and
\citet{07Irikur.gen}.  In an experimental study carried out by
\citet{03SaMcTh.SiO}, Dunham coefficients and BOB correction terms
were determined for the SiS ground electronic state (X $^1\Sigma^+$)
using Fourier Transform Microwave (FTM) spectroscopy. These
coefficients were used as spectroscopic constants by
\citet{16BaCoxx.partfunc} to calculate their partition function values
listed in Table~\ref{t:pf}, which are in particularly good agreement
with our (direct summation of energy level) values at lower
temperatures. Our partition functions for all 12 isotopologues on a 1
K grid up to $T = 5000$ K are provided in the supplementary data.

\begin{table}
\caption{Comparison of our partition function for $^{28}$Si$^{32}$S  with 
the values given in CDMS \citep{cdms} and by \citet{16BaCoxx.partfunc}
as a function of temperature, $T$.}
\label{t:pf}
\begin{tabular}{crrr}
\hline\hline
\multicolumn{1}{c}{$T$(K)}		&
      \multicolumn{1}{c}{This Work}	&
\multicolumn{1}{c}{CDMS} & \multicolumn{1}{c}{\citet{16BaCoxx.partfunc}}
\\
\hline
3.0   &    7.22963 &           &  7.22968   \\
9.375 &   21.8566  &   21.8566 &            \\
18.75 &   43.3764  &  43.3765  &            \\
20.0  &  46.2461   &           &  46.2464   \\
37.5  &  86.4222   & 86.4221   &            \\
75.0  & 172.530    &172.529    &            \\
130.0 & 298.935    &           & 298.939    \\
150.0 & 345.078    &345.077    &            \\
225.0 & 521.644    &521.643    &            \\
300.0 & 709.671    &709.670    &            \\
500.0 & 1303.91    &1303.91    & 1303.94    \\
1000.0& 3519.30    &3519.25    &            \\
3000.0& 23769.6    &           &23774.3     \\
8000.0& 156220     &           &178366      \\
\hline\hline                                
\end{tabular}                               
\end{table}

For ease of use the partition functions are also
was fitted to the functional form proposed by \citet{jt263}
\begin{equation}\label{eq:fit1d}
\log_{10}Q(T) = \sum\limits^{8}_{n=0} a_n (\log_{10}T)^n.
\end{equation}
The fitted expansion parameters for each isotopologue are given in the supplementary material.
These parameters reproduce the temperature dependence of
partition function of SiS with a relative root-mean-square error of  0.0076
up to  $T$ = 5000~K which is the maximum temperature for 
which our line list is recommended.

\subsection{Transition frequencies}
We initially computed all rotation-vibrational transitions
in the ground electronic state which satisfy the selection rule
$\Delta J = \pm 1$, with these transitions
occurring between states as high as $v = 43$ and
$J = 257$. There are around 330~000  transitions in the case of $^{28}$Si$^{32}$S.
However, given concerns with the numerical stability of the intensity of
higher overtone transitions \citep{highv}, we chose to eliminate all
transitions with $\Delta v \geq 6$. This reduces each line list to less
than 100~000 transitions.


Table~\ref{t:trans} compares our computed transition frequencies with
a selection of measured frequencies covering a range of vibrational
and rotational states for $^{28}$Si$^{32}$S. The agreement is excellent;
essentially to within the experimental uncertainty of 0.001 \cm\ quoted by
\citet{90BiJoxx.SiS} for their measurements. Table~\ref{t:trans2} gives
a similar comparison, albeit for a reduced range of vibrational states,
for three isotopologues of SiS. Again agreement is within experimental error.
These comparisons provide confidence about the accuracy of the
lines positions in the line list.

\begin{table}
\caption{Comparison of our predicted (Calc) transition frequencies (\cm) with experimentally obtained (Obs) values 
by \citet{90BiJoxx.SiS} for the parent isotopologue $^{28}$Si$^{32}$S.}
\label{t:trans}
\begin{tabular}{rrccccr}
\hline\hline
$J^\prime$&$J^{\prime\prime}$&$v^\prime$&$v^{\prime\prime}$&
      \multicolumn{1}{c}{Obs}	&
\multicolumn{1}{c}{Calc} & \multicolumn{1}{c}{Obs$-$Calc}
\\
\hline
88 &  89 &   1 &  0&  679.5896 &  679.5905&  -0.0009  \\
6  &   5 &   1 &  0&  748.0476 &  748.0478&  -0.0002  \\
99 &  100&   2 &  1&  665.2391 &  665.2403&  -0.0012  \\
116&  115&   2 &  1&  787.9384 &  787.9391&  -0.0007  \\
6  &  7  &   3 &  2&  729.8961 &  729.8963&  -0.0002  \\
69 &  68 &   3 &  2&  768.1500 &  768.1498&   0.0002  \\
42 &  43 &   4 &  3&  700.7459 &  700.7453&   0.0006  \\
34 &  33 &   4 &  3&  747.5076 &  747.5078&  -0.0001  \\
89 &  90 &   5 &  4&  659.1780 &  659.1773&   0.0008  \\
24 &  23 &   5 &  4&  737.2133 &  737.2134&  -0.0001  \\
17 &  18 &   6 &  5&  707.6352 &  707.6353&  -0.0001  \\
127&  126&   6 &  5&  768.0828 &  768.0801&   0.0027  \\
20 &  21 &   7 &  6&  700.6245 &  700.6236&   0.0008  \\
74 &  73 &   7 &  6&  748.5679 &  748.5681&  -0.0002  \\
53 & 54  & 8 & 7& 672.7660 & 672.7658&  0.0002    \\
90 & 89  & 8 & 7& 748.4143 & 748.4110&  0.0033    \\
25 & 26  & 9 & 8& 687.2477 & 687.2482& -0.0005    \\
8  & 7   & 9 & 8& 707.8742 & 707.8747& -0.0005    \\
29 & 30  &10 & 9& 679.5703 & 679.5700&  0.0003    \\
19 & 20  &10 & 9& 686.0696 & 686.0688&  0.0008    \\
\hline \hline                                           
\end{tabular}
\end{table}

\begin{table}
\caption{Comparison of our predicted (Calc) transition frequencies (\cm) with experimentally obtained (Obs) values 
by \citet{90BiJoxx.SiS} for isotopically subsituted SiS.}
\label{t:trans2}
\begin{tabular}{lrrccccr}
\hline\hline
&$J^\prime$&$J^{\prime\prime}$&$v^\prime$&$v^{\prime\prime}$&
      \multicolumn{1}{c}{Obs}	&
\multicolumn{1}{c}{Calc} & \multicolumn{1}{c}{Obs$-$Calc}
\\
\hline
$^{28}$Si$^{34}$S   
&82 & 83&  1&  0 & 676.1871& 676.1873 &-0.0002  \\
&24 & 23&  1&  0 & 747.5463& 747.5470 &-0.0007  \\
&63 & 64&  2&  1 & 686.2286& 686.2290 &-0.0004  \\
&6  & 5 &  2&  1 & 732.7039& 732.7050 &-0.0010  \\
&36 & 37&  3&  2 & 700.8034& 700.8026 & 0.0008  \\
&88 & 87&  3&  2 & 763.9691& 763.9694 &-0.0003  \\
&49 & 50&  4&  3 & 686.8249& 686.8243 & 0.0006  \\
&26 & 25&  4&  3 & 733.3080& 733.3081 &-0.0001  \\
                                 \\
$^{30}$Si$^{32}$S     
&106& 107 &1 & 0 & 658.7018& 658.7018&  0.0000 \\
&39 & 38  &1 & 0 & 758.5463& 758.5460&  0.0003 \\
&4  & 5   &2 & 1 & 729.5870& 729.5884& -0.0014 \\
&2  & 1   &2 & 1 & 733.7504& 733.7495&  0.0009 \\
&40 &41   &3 & 2 & 701.0780& 701.0768&  0.0012 \\
&90 &89   &3 & 2 & 768.1860& 768.1862& -0.0002 \\
&54 &55   &4 & 3 & 686.1028& 686.1023&  0.0005 \\
&26 &25   &4 & 3 & 736.6676& 736.6676&  0.0000 \\
                                   \\
$^{30}$Si$^{32}$S                   
&8 &  9  & 1&  0 & 725.8659 &725.8648&  0.0011    \\
&67& 66  & 1&  0 & 763.7698 &763.7715& -0.0017    \\
&34& 35  & 2&  1 & 704.2666 &704.2667& -0.0001    \\
&95& 94  & 2&  1 & 768.0828 &768.0861& -0.0034    \\
&21& 22  & 3&  2 & 707.9016 &707.9028& -0.0012    \\
&36& 35  & 3&  2 & 740.2007 &740.1993&  0.0014    \\
&74& 75  & 4&  3 & 665.6780 &665.6789& -0.0009    \\
&40& 39  & 4&  3 & 736.9890 &736.9893& -0.0003    \\
  \hline\hline
\end{tabular}
\end{table}

\begin{figure}
\centering
\includegraphics[scale=0.6]{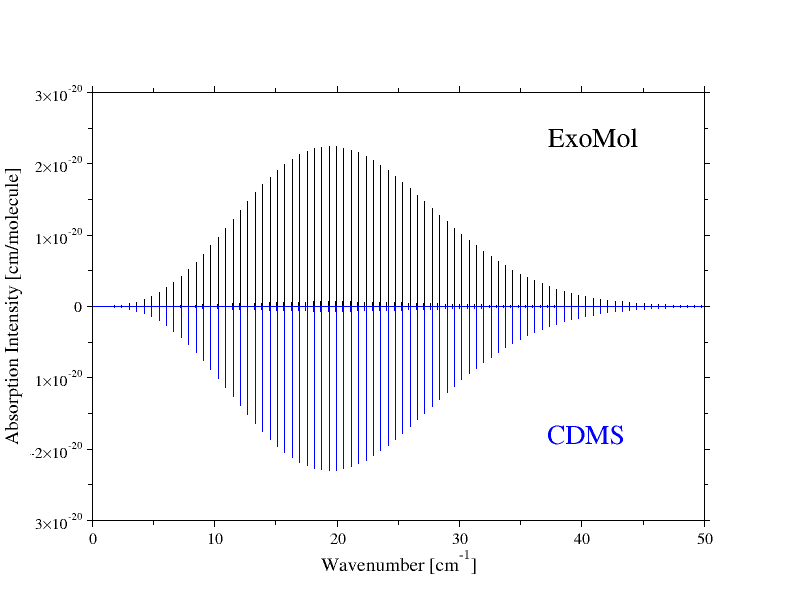}
\caption{Comparison of our (ExoMol) pure rotational absorption spectrum of $^{28}$Si$^{32}$S at $T= 300$~K in
comparison with that given by CDMS.}
\label{f:rot}
\end{figure}

\begin{figure}
\centering
\includegraphics[scale=0.6]{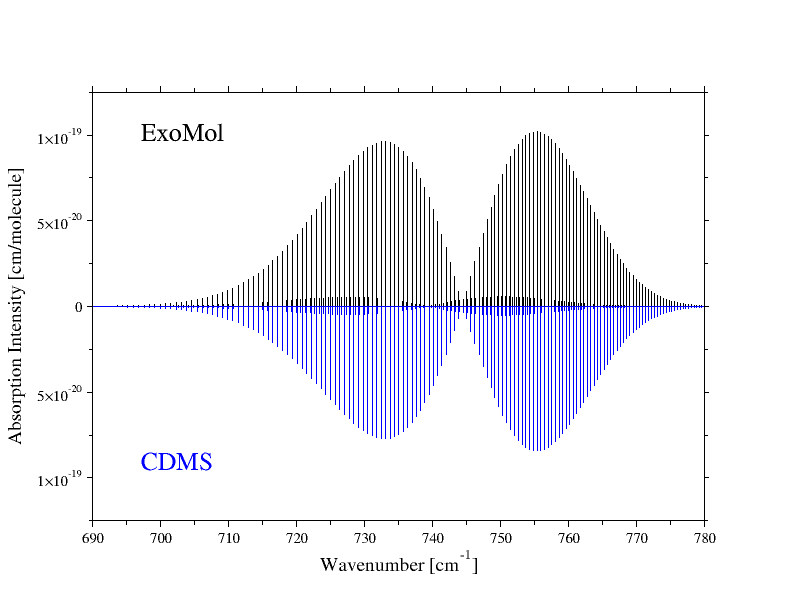}
\caption{Comparison of our (ExoMol) absorption spectrum of $^{28}$Si$^{32}$S 
vibrational fundamental at $T= 300$~K 
with that given by CDMS.}
\label{f:vib}
\end{figure}

\subsection{Comparisons of spectra}

In order to test the quality of our theoretical line list, we present
comparisons with previous works where possible. For SiS the CDMS catalogue
\citep{cdms}, rather unusually, contains both pure rotational and
vibration-rotation spectra for several isotopologues of SiS.
Figures \ref{f:rot} and \ref{f:vib} compare our predictions for
$^{28}$Si$^{32}$S with those of CMDS. For the pure rotational
spectrum, Fig. \ref{f:rot}, the agreement is excellent. CDMS is
carefully designed to be highly accurate for such long wavelength
spectra and anyone wishing to study low-temperature rotational
transitions of SiS is advised to start from the data in CDMS. The
comparison for the vibrational fundamental, Fig. \ref{f:vib}, is less
good. In particular our spectrum is significantly stronger than the
one given by CDMS due to our \ai\ transition dipole value (0.14 D)
 being slightly higher than the semi-empirical estimate (0.13 D) provided by \citet{87PiTiCh.SiS}
for the $v=1 -  0$ transition. 
In this case we expect our results to be more
reliable since CDMS uses a rather simple treatment of the transition
dipole whereas our calculation is based on the use of a
state-of-the-art dipole moment function.

\begin{figure}
\includegraphics[scale=0.6]{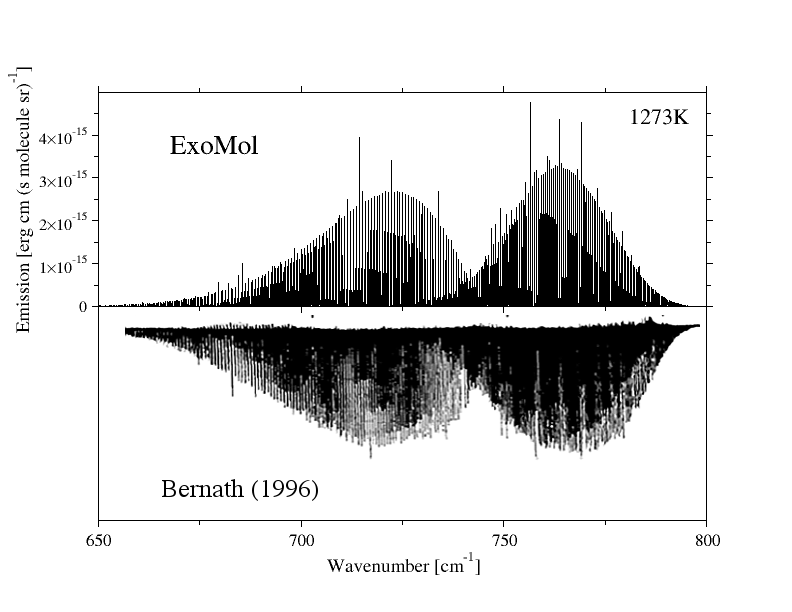}
\caption{Comparison of our (ExoMol) emission spectrum of $^{28}$Si$^{32}$S at 1000 $^\circ$C with
the laboratory spectrum given by \protect{\citet{96Bernat}}.} 
\label{f:emitb}
\end{figure}

\begin{figure}
\includegraphics[scale=0.6]{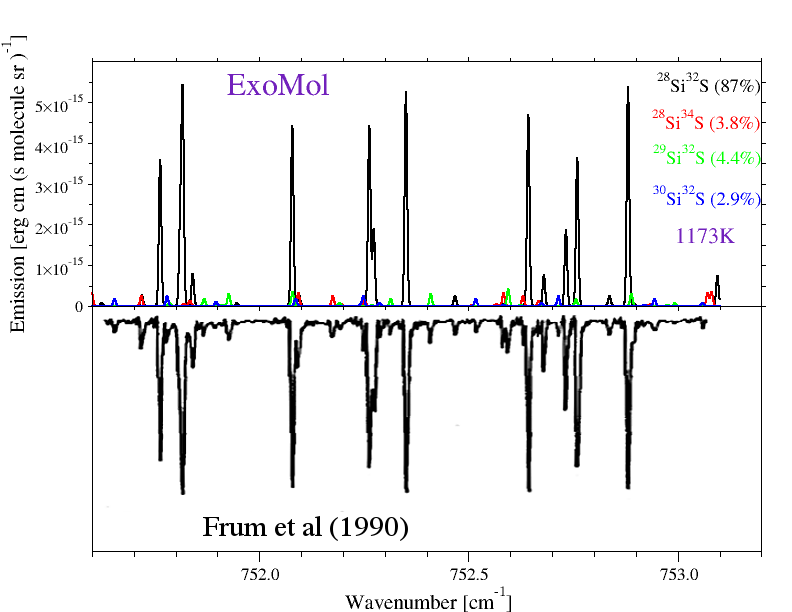}
\caption{Comparison of our (ExoMol) emission spectrum of $^{28}$Si$^{32}$S at 900 $^\circ$C with
the laboratory spectrum given by \protect{\citet{90FrEnBe.SiS}}. Contributions
from the different isotopologues, assumed to be in natural abundance,
are highlighted in our spectrum.} 
\label{f:emitf}
\end{figure}

There are very limited data available on hot SiS spectra. An exception
is the 13~$\mu$m region; an overview emission spectrum for this region was presented by
\citet{96Bernat} based on the measurements of \citet{90FrEnBe.SiS}.
Figure \ref{f:emitb} compares our predictions with this experiment.
Given the relative crude nature of the observed spectrum, for which no
absolute intensities are available, agreement must be regarded as
satisfactory. In particular P and R branch with the vibrational band
$v=1\rightarrow 0$, $v=2\rightarrow 1$, $v=3\rightarrow 2$ and
$v=4\rightarrow 3$, in order of decreasing intensity, are clearly
visible. \citet{96Bernat} notes similar features in his spectrum. We
note that at higher resolution there are observable contributions from several
isotopologues, as shown in Figure \ref{f:emitf}.

\subsection{Overview}

In accordance with ExoMol format \citep{jt631}, the line lists are
presented as two files: a states file and a transitions file.
Tables~\ref{table:states} and \ref{table:trans} give brief abstracts
of the $^{28}$Si$^{32}$S states and transitions files, respectively.
These files can be combined with the partition function, which is also
provided in the database, to give the desired spectrum at a given
temperature. These files are made available for all
12 isotopologues considered at 
\url{ftp://cdsarc.u-strasbg.fr/pub/cats/J/MNRAS/xxx/yy}, or
\url{http://cdsarc.u-strasbg.fr/viz-bin/qcat?J/MNRAS//xxx/yy} as well
as the ExoMol website, \url{www.exomol.com}.

\begin{table*}
\centering
\caption{Extract from the states file of the $^{28}$Si$^{32}$S line list.}
\label{table:states}
{\tt
 \begin{tabular}{rcrrr}
 \hline
 \hline
$n$	&	$\tilde{E}$	&	$g_i$	&	$J$	&	$v$	\\
\hline
1&	0.000000	&1&	0&	0      \\
2&	0.605581	&3&	1&	0      \\
3&	1.816740	&5&	2&	0      \\
4&	3.633466	&7&	3&	0      \\
5&	6.055745	&9&	4&	0      \\
6&	9.083558	&11&	5&	0     \\
\hline
\hline
\end{tabular}
}
\mbox{}\\
{\flushleft
$n$:   State counting number.     \\
$\tilde{E}$: State energy in \cm. \\
$g_i$:  Total statistical weight, equal to ${g_{\rm ns}(2J + 1)}$.     \\
$J$: Total angular momentum.\\
$v$:   State vibrational quantum number. \\
}
\end{table*}

\begin{table}
\caption{Extract from the transitions file of the $^{28}$Si$^{32}$S line list.}
\label{table:trans}
{\tt
\begin{tabular}{rrrr}
\hline\hline
\multicolumn{1}{c}{$f$}	&	\multicolumn{1}{c}{$i$}	&
\multicolumn{1}{c}{$A_{fi}$ (s$^{-1}$)}	&\multicolumn{1}{c}{$\tilde{\nu}_{fi}$}	\\
\hline
9972&	9971&	6.9703E-08&	0.484833\\
9854&	9853&	7.1086E-08&	0.486543\\
9713&	9712&	7.2306E-08&	0.489422\\
9552&	9551&	7.3278E-08&	0.492438\\
9374&	9373&	7.4098E-08&	0.495470\\
9179&	9178&	7.4809E-08&	0.498500\\
\hline\hline
\end{tabular}
}

\noindent
 $f$: Upper  state counting number;\\
$i$:  Lower  state counting number; \\
$A_{fi}$:  Einstein-A coefficient in s$^{-1}$; \\
$\tilde{\nu}_{fi}$: transition wavenumber in \cm.\\
\end{table}

\begin{figure}
\centering
\includegraphics[scale=0.6]{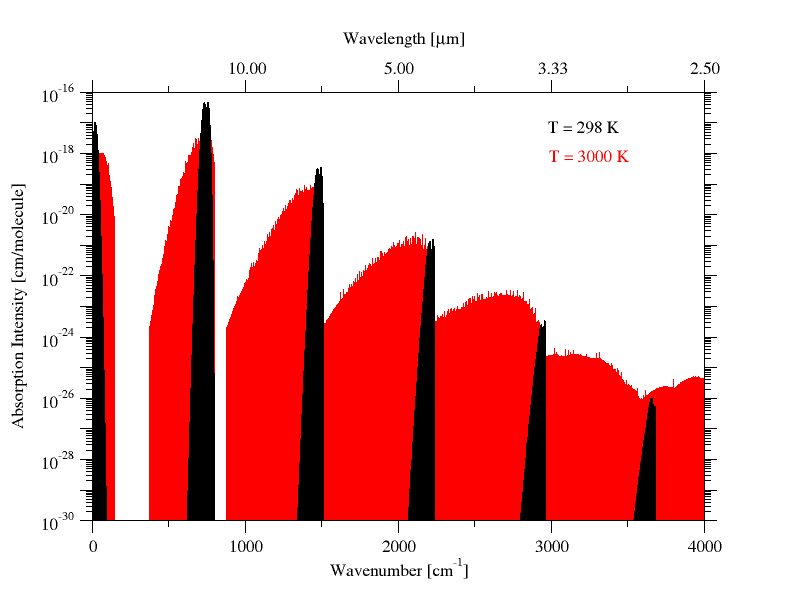}
\caption{Temperature-dependent absorption spectrum of $^{28}$Si$^{32}$S.}
\label{f:T}
\end{figure}

Figure~\ref{f:T} presents an overview of the SiS absorption spectrum
as a function of temperature.

\section{Conclusion}
\label{s:concl}

Accurate and complete line lists for 12 isotopologues of SiS
are presented. The line lists, which we call UCTY,  use potential energy curves based on the
highly accurate study of \citet{92CoHaxx.SiS} and newly computed dipole moment functions.
They represent the first complete line lists for these systems.

The detection of many hot rocky planets, so called lava planets, has significantly
increased the number of small molecules whose spectra may be important in exoplanet
atmospheres \citep{jt693}; SiS is one of these species. We hope that line lists
such as the ones presented here will aid the characterisation of exoplanetary
atmospheres by planned observational missions such as 
ARIEL \citep{jt717} and Twinkle \citep{jt659}.

\section*{Acknowledgements}

This work was supported by the UK Science and Technology Research
Council (STFC) grant No. ST/M001334/1 and the COST action MOLIM No. CM1405.
This work made extensive use of UCL's Legion high performance
computing facility.

\bibliographystyle{mnras}

\label{lastpage}
\end{document}